\documentclass[useAMS,usenatbib,twcolumn,usegraphicx]{mn2e}

\newcommand{\gsim}{\, \raisebox{-0.8ex}{$\stackrel{\textstyle >}{\sim}$ }}
\newcommand{\lsim}{\, \, \raisebox{-0.8ex}{$\stackrel{\textstyle <}{\sim}$ }}

\newcommand{\beq}{\begin{equation}}
\newcommand{\eeq}{\end{equation}}
\newcommand{\beqar}{\begin{eqnarray}}
\newcommand{\eeqar}{\end{eqnarray}}

\title[Effect of superfluidity on NS crustal oscillations]
{Effect of superfluidity on neutron star crustal oscillations}
\author[H. Sotani, K. Nakazato, K. Iida, \& K. Oyamatsu]
{Hajime Sotani$^1$ \thanks{E-mail:hajime.sotani@nao.ac.jp},
Ken'ichiro Nakazato$^2$,
Kei Iida$^3$, and
Kazuhiro Oyamatsu$^4$
\\
$^1$Division of Theoretical Astronomy, National Astronomical Observatory of Japan, 
2-21-1 Osawa, Mitaka, Tokyo 181-8588, Japan\\
$^2$Faculty of Science \& Technology, Tokyo University of Science, 2641 Yamazaki, Noda, Chiba 278-8510, Japan\\
$^3$Department of Natural Science, Kochi University, 2-5-1 Akebono-cho, Kochi 780-8520, Japan\\
$^4$Department of Human Informatics, Aichi Shukutoku University, 9 Katahira, Nagakute, Aichi 480-1197, Japan}
\begin{document}

\maketitle

\label{firstpage}

\begin{abstract}
We consider how superfluidity of dripped neutrons in the crust of a 
neutron star affects the frequencies of the crust's fundamental torsional 
oscillations.  A nonnegligible superfluid part of dripped neutrons, which 
do not comove with nuclei, act to reduce the enthalpy density and thus 
enhance the oscillation frequencies.  By assuming that the quasi-periodic 
oscillations observed in giant flares of soft gamma repeaters arise
from the fundamental torsional oscillations and that the mass and 
radius of the neutron star is in the range of $1.4\le M/M_\odot\le 1.8$ 
and 10 km $\le R \le$ 14 km, we constrain the density derivative of the 
symmetry energy as 100 MeV $\lsim L \lsim$ 130 MeV, which is 
far severer than the previous one, $L\gsim 50$ MeV, derived by ignoring
the superfluidity.
\end{abstract}

\begin{keywords}
relativity -- stars: neutron -- stars: oscillations  -- equation of state
\end{keywords}


The structure of neutron stars remains uncertain mainly because 
theoretical extrapolations from empirically known properties of atomic 
nuclei are required to describe the equation of state (EOS) of high density 
matter inside the star.  Consequently, the theoretical neutron star 
mass ($M$) - radius ($R$) relation is still model dependent, whereas 
the presence of a neutron star that has mass of about $2M_\odot$ has 
recently been confirmed \citep{2Msun} and plays a role in ruling out some 
EOS models.  It is generally accepted that under 
an ionic ocean near the star's surface and above a fluid core, 
there exists a crustal region, where nuclei form a bcc Coulomb 
lattice in a roughly uniform electron sea and at a density 
of about $4\times 10^{11}$ g cm$^{-3}$, neutrons drip out of the
nuclei.  The resultant neutron gas is  a superfluid, as long as the temperature
is below the critical temperature, and presumably 
relevant to pulsar glitches \citep{Sauls}.  
In fact, most neutron stars observed
are expected to be cool enough to have a crust
containing a neutron superfluid.
Even at zero temperature, however, 
a part of the neutron gas comoves non-dissipatively with protons in the nuclei via 
Bragg scattering off the lattice \citep{Chamel2012}.  Crustal torsional 
oscillations, if occurring in stars, would be controlled by
the enthalpy density of the constituents that comove with 
the protons \citep{vanhorn90}.  In this article, we thus address how neutron 
superfluidity affects the oscillation frequencies in a manner that 
depends on the still uncertain EOS of neutron-rich nuclear matter. 
We remark that there are
earlier publications that examine the influence of neutron
superfluidity on the crustal torsional oscillations for a
specific model of the crust EOS \citep{AGS2009,SA2009,PA2012}.
As we shall see, the effects of neutron
superfluidity as analyzed here are consistent with the
results obtained in these publications.

It is not long ago that oscillations of neutron stars were observed 
in the form of quasi-periodic oscillations (QPOs) in giant flares from 
soft gamma repeaters (SGRs), which are considered to be magnetars, 
i.e., neutron stars with surface magnetic fields of order $10^{14-15}$ G.
This causes neutron star asteroseismology to become of practical 
significance in studying the properties of matter in the star 
(e.g., \cite{AK1996,Sotani2001,Sotani2004,SYMT2011}). 
So far, 
three giant flares have been detected from SGR 0526-66, SGR 1900+14, and 
SGR 1806-20.  The frequencies of the QPOs discovered through the timing 
analysis of the afterglow are in the range from tens Hz up to a few kHz 
\citep{WS2006}.  Many theoretical attempts to explain the observed 
QPO frequencies have been done in terms of the torsional oscillations 
in the crustal region and/or the magnetic oscillations (e.g.,  
\cite{Levin2006,Lee2007,SA2007,Sotani2007,Sotani2008a,Sotani2008b,Sotani2009}).
The important conclusion is that either the torsional oscillations or 
magnetic oscillations dominate the excited oscillations in magnetized 
neutron stars, depending on the magnetic field strength  
\citep{CK2011,GCFMS2011,GCFMS2012a} and configurations \citep{GCFMS2012b}. 
In the absence of the observational information about
the magnetic structure of magnetars except the inferred
field strengths from the observed spindown \citep{K1998,H1999},
we assume for simplicity that the QPOs observed 
in giant flares are associated with the crustal oscillations.  Under 
this assumption, one can probe the properties of nuclear matter in the crust
such as the density dependence of the symmetry energy and neutron 
superfluidity \citep{SW2009,Sotani2011b,GNHL2011,SNIO2012}.  Once the density 
dependence of the symmetry energy is determined, one can address 
how large the region in which pasta nuclei (rods, slabs, etc.) 
\citep{LRP1993,O1993} occur would be, if any \citep{OI2007}.


To quantify the density dependence of the symmetry energy, we use the 
expansion of the bulk energy per nucleon near the saturation point of 
symmetric matter at zero temperature \citep{L1981}:
\begin{equation}
  w=w_0 + \frac{K_0}{18n_0^2}(n-n_0)^2+\left[S_0+\frac{L}{3n_0}(n-n_0)\right]
\alpha^2, \label{eq:bulk-en}
\end{equation}
where $n$ is the nucleon density, $\alpha$ is the neutron excess,
$w_0$, $n_0$, and $K_0$ denote the saturation energy, saturation density, and 
incompressibility of symmetric nuclear matter, respectively, and $S_0$ and $L$ 
are the parameters characterizing the symmetry energy coefficient $S(n)$, i.e.,
$S_0\equiv S(n_0)$ and $L=3n_0(dS/dn)_{n=n_0}$.  The parameters $w_0$, $n_0$, 
and $S_0$ can be well determined from the empirical masses and 
radii of stable nuclei \citep{OI2003}, while the parameters $L$ and $K_0$ are
relatively uncertain.  In fact, two of us (K.O. \& K.I.) constructed the 
model for the EOS of nuclear matter in such a way as to reproduce
Eq.\ (\ref{eq:bulk-en}) in the limit of $n\to n_0$ and $\alpha\to 0$. Then, 
the optimal density distribution of stable nuclei was obtained from
the EOS model within a simplified version of the extended Thomas-Fermi theory.
Finally, for given $y(\equiv -K_0S_0/3n_0L)$ and $K_0$, the values of 
$w_0$, $n_0$, and $S_0$ were obtained by fitting the charge number, mass 
excess, and charge radius calculated from the optimal density distribution to
the empirical ones.

To describe matter in the crust, the Thomas-Fermi model was generalized
by adding dripped neutrons, a neutralizing background of electrons,
and the lattice energy within a Wigner-Seitz approximation \citep{OI2007}. 
This allows one to obtain the equilibrium nuclear shape and size as well as 
the crust EOS for various sets of $y$ and $K_0$.  
Here, as in \cite{OI2007} and \cite{SNIO2012}, we adopt the parameter range 
$0< L < 160$ MeV, $180 \le K_0 \le 360$ MeV, and $y< -200$ MeV fm$^3$, which 
can reproduce the mass and radius data for stable nuclei and effectively cover 
even extreme cases \citep{OI2003} 
(see Table I in \cite{SNIO2012} for the adopted EOS parameters).

 We turn to calculations of the eigenfrequencies of the crustal
torsional oscillations.  We first prepare the equilibrium configuration 
of the non-rotating neutron stars, which is the 
spherically symmetric solution of the Tolman-Oppenheimer-Volkoff (TOV) 
equations.  
The corresponding solution is described in the following metric 
with the spherical polar coordinate:
\begin{equation}
  ds^2 = -{\rm e}^{2\Phi(r)} dt^2 + {\rm e}^{2\Lambda(r)} dr^2 + r^2 (d\theta^2 + \sin^2\theta\,d\phi^2).
\end{equation} 
In constructing a neutron star from its center, one needs the EOS for 
matter in the core as well as the crust EOS.  Although the core EOS 
dominates $M$ and $R$, even the constituents of the core 
remain uncertain.  We thus concentrate on the crustal region without 
specifying the core EOS.  In doing so, for various sets of $M$ and $R$ and
various models for the crust EOS, we integrate the TOV equations inward 
from the star's surface, as in \cite{IS1997} and \cite{SNIO2012}.  Hereafter,
we shall consider typical values of $M$ and $R$, namely, 
$1.4\le M/M_\odot\le 1.8$ and $10$ km $\le R\le$ 14 km.  Note that such
choice of $M$ and $R$, which is rather arbitrary, encapsulates uncertainties
of the core EOS.

In the presence of torsional oscillations, the elasticity of the 
crust works as a restoring force.  This elasticity is characterized by 
the shear modulus $\mu$, which is another ingredient of the
calculations of the oscillation frequencies. 
By assuming that spherical nuclei with charge 
$Ze$ form a bcc Coulomb crystal with number density $n_i$, one can 
approximately describe the shear modulus averaged over all directions in the 
limit of zero temperature as $\mu = 0.1194 n_i (Ze)^2/a$, where $a\equiv 
(3/4\pi n_i)^{1/3}$ is the Wigner-Seitz radius 
\citep{OI1990,SHOII1991}. 
As shown in \cite{SNIO2012}, this modulus, which depends on $L$ 
mainly through the $L$ dependence of $Z$ \citep{OI2007}, controls the $L$ 
dependence of the oscillation frequencies.
Note that pasta nuclei, if present, would contribute to the restoring 
force in a manner that depends on the shapes \citep{PP1998}.  However, the 
values of $L$ that will be constrained from the present analysis suggest that 
the region where pasta nuclei occur is highly limited \citep{OI2007}.  Thus,
as in \cite{GNHL2011} and \cite{SNIO2012}, we simply consider the torsional 
oscillations to be confined within the region of spherical nuclei.


Calculations of the frequencies of the torsional modes
on the spherical equilibrium configuration of the crust as
obtained above can be performed with high accuracy by using
the relativistic Cowling approximation,
namely, neglecting the metric perturbations. 
This is due to the incompressibility of the modes.
With the $\phi$-component of the Lagrangian displacement vector
of a matter element given by $\xi^\phi={\cal Y}(t,r)
\partial_\theta P_\ell/\sin\theta$, where $P_\ell$ is
the $\ell$-th order Legendre polynomial,
the perturbation equation 
governing the torsional oscillations can be derived from the linearized 
equation of motion as \citep{ST1983}
\begin{eqnarray}
 \mu r^2{\cal Y}'' &+& [(4+r\Phi'-r\Lambda')\mu r+\mu'r^2]{\cal Y}'  \nonumber \\
   &+& [r^2H\omega^2{\rm e}^{-2\Phi}-(\ell+2)(\ell-1)\mu]{\rm e}^{2\Lambda}{\cal Y} = 0.
 \label{eq:perturbation}
\end{eqnarray}
Here, we assume ${\cal Y}={\rm e}^{{\rm i}\omega t}{\cal Y}(r)$, the 
prime denotes the derivative with respect to $r$, and $H$ is the enthalpy 
density defined as $H\equiv \epsilon+p$ with
the pressure $p$ and the energy density $\epsilon$.  We remark that at zero 
temperature, the baryon chemical potential $\mu_b$ can be expressed as 
$\mu_b=H/n_b$ with the baryon density $n_b$.  The boundary 
conditions to be imposed in determining the frequency $\omega$ of 
the modes are the zero-traction condition at the density where 
spherical nuclei ceases to be present in the deepest region of the crust and 
the zero-torque condition at the star's surface \citep{ST1983,Sotani2007}.

     We now consider the effect of neutron superfluidity. It is generally 
accepted that at a density of about $4\times 10^{11}$ g cm$^{-3}$,
neutrons start to be dripped out of nuclei and form a superfluid.  Even at zero
temperature, however, a component that undergoes Bragg scattering from
the bcc lattice of the nuclei and thus move non-dissipatively with the nuclei is still 
present.  According to recent band calculations beyond the Wigner-Seitz
approximation by \cite{Chamel2012}, the superfluid density,
which we define as the density of neutrons that are not locked to the motion of 
protons in the nuclei,
depends sensitively on the baryon density above neutron drip.  Since the 
crustal torsional oscillations are transverse, 
only neutrons that are responsible for the
superfluid density behave
independently of the associated displacements of the nuclei \citep{PCR2010}.
However, the enthalpy density $H$ in Eq.\ (\ref{eq:perturbation}) fully 
contains the superfluid mass density \citep{IB-II}.  We thus subtract the 
superfluid mass density from $H$ and obtain the effective enthalpy density 
$\tilde{H}=H\times(A-N_s)/A$, where $A$ denotes the total nucleon 
number in a Wigner-Seitz cell, and $N_s$ denotes the number of neutrons 
in a Wigner-Seitz cell that do not comove with the nucleus.
Finally, by 
substituting $\tilde{H}$ for $H$ in Eq.\ (\ref{eq:perturbation}), one can 
determine the frequencies of the torsional oscillations in a manner that 
depends on $N_s$.  Hereafter we will assume that $N_s$ comes entirely from
the gas of dripped neutrons.  In fact, although it could come partially 
from neutrons inside a nucleus, the dripped neutron gas dominates $A$ in 
the entire density region except just above neutron drip.

\begin{figure}
\begin{center}
\includegraphics[scale=0.5]{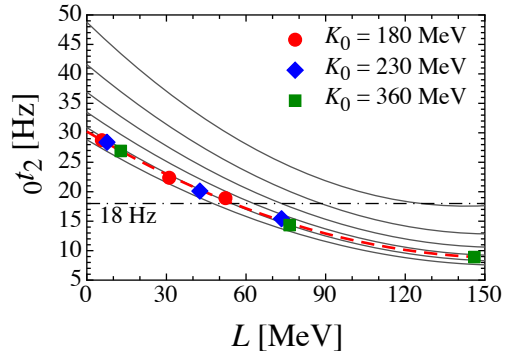} 
\end{center}
\caption{
(Color online) Frequencies of the $\ell=2$ fundamental torsional oscillations
of a star having $M=1.8M_\odot$ and $R=14$ km,  ${}_0t_2$, 
plotted as a function of $L$. 
The six solid lines from bottom to top correspond to the cases of 
$N_s/N_d=0$, 0.2, 0.4, 0.6, 0.8,
and 1, while the broken line with symbols shows the result 
from the values of $N_s/N_d$ derived by Chamel (2012). The 
horizontal dot-dashed line denotes the lowest QPO frequency observed from SGR 
1806-20.
}
\label{fig:0t2Lrate}
\end{figure}

We show the results for the frequency of the fundamental $\ell=2$ 
torsional oscillations, ${}_0t_2$, calculated for various values of 
$N_s/N_d$ with the number $N_d$ of dripped neutrons in a Wigner-Seitz cell. 
We remark that $N_d-N_s$ corresponds to the number of dripped 
neutrons bound to the nuclei.
First, for simplicity, we set the ratio $N_s/N_d$ constant throughout the 
density region above neutron drip.  At given $M$, $R$, and $N_s/N_d$, 
we calculate ${}_0t_2$ for nine sets of the EOS parameters, as in 
\cite{SNIO2012}.  Since the calculated ${}_0t_2$ has a negligible 
dependence on $K_0$, we construct a fitting formula as 
${}_0t_2=c_0-c_1L+c_2L^2$, where $c_0$, $c_1$, and $c_2$ are positive and
adjusted in such a way as to well reproduce the calculations for the nine 
cases.  The formulas obtained for $N_s/N_d=0$, 0.2, 0.4, 0.6, 0.8, 
1, $M=1.8M_\odot$, and $R=14$ km are plotted in Fig.\ \ref{fig:0t2Lrate}. 
From this figure, one can observe that, due to neutron superfluidity,
which acts to enhance the shear speed defined as $v_s=(\mu/\tilde{H})^{1/2}$, ${}_0t_2$
becomes up to $66-82\%$ larger for $L\le 76.4$ MeV and up to
$131\%$ larger for $L=146.1$ MeV 
than that without neutron superfluidity.
Additionally, in this figure we plot
the result for the stellar model with the values of $N_s/N_d$ derived 
from band calculations in \cite{Chamel2012} as ``$n_n^{\rm c}/n_n^{\rm f}$" in Table I.
Since these values are 
only of order 10--$30\%$ at $n_b\sim0.01$--$0.4n_0$, the increase in ${}_0t_2$ 
is relativity small and close to the result obtained at $N_s/N_d=0.2$.
Such increase in ${}_0t_2$ is consistent with
the results of \cite{AGS2009,SA2009,PA2012}.

This increase of ${}_0t_2$ due to neutron superfluidity
would make the constraint of $L$ severer if the QPO frequencies 
observed in SGRs come from the torsional oscillations. In this case, 
the lowest frequency in the observed QPOs should be equal to or larger than 
the predicted ${}_0t_2$, because ${}_0t_2$ is the lowest frequency 
among various eigenfrequencies of the torsional oscillations.  It is
also important to note the tendency that, as $M$ and $R$ become larger, 
${}_0t_2$ becomes smaller \citep{SNIO2012}.  Consequently, ${}_0t_2$
obtained for $M=1.8M_\odot$ and $R=14$ km would determine the lower 
limit of the allowed $L$ from the observed QPO frequencies, as 
long as we confine the stellar models within $1.4\le M/M_\odot\le 1.8$ and 
$10$ km $\le R\le$ 14 km.

\begin{figure}
\begin{center}
\includegraphics[scale=0.5]{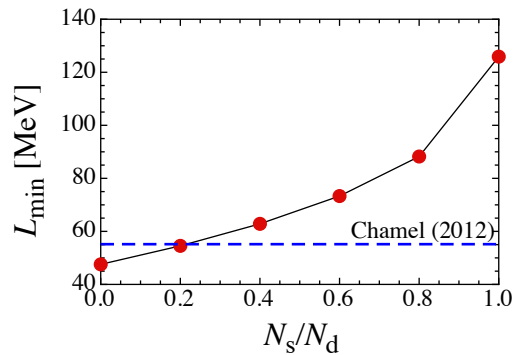} 
\end{center}
\caption{
(Color online) $L_{\rm min}$ as a function of $N_s/N_d$.
The horizontal broken line corresponds to the 
result from the band calculations of $N_s/N_d$ by Chamel.
}
\label{fig:Lmin}
\end{figure}

    This lower limit of $L$, hereafter referred to as $L_{\rm min}$, 
can be obtained from the intersections between the horizontal dot-dashed 
line of 18 Hz and the lines of ${}_0t_2$ in Fig.\ \ref{fig:0t2Lrate}.  For constant $N_s/N_d$, the 
resultant $L_{\rm min}$ is shown in  Fig.\ \ref{fig:Lmin}.  One can
observe that the value of $L_{\rm min}$, which is 47.6 MeV in the absence
of neutron superfluidity (i.e., $N_s/N_d=0$), can be as 
large as 125.9 MeV in its presence (i.e., $0<N_s/N_d\leq1$).
In addition, we exhibit $L_{\rm min}=55.2$ MeV, the result from the broken
line in Fig.\ \ref{fig:0t2Lrate}.  This $L_{\rm min}$, which comes from 
the realistic band calculations of $N_s/N_d$ in \cite{Chamel2012}, 
is expected to give a reliable constraint on $L$.

\begin{figure}
\begin{center}
\includegraphics[scale=0.5]{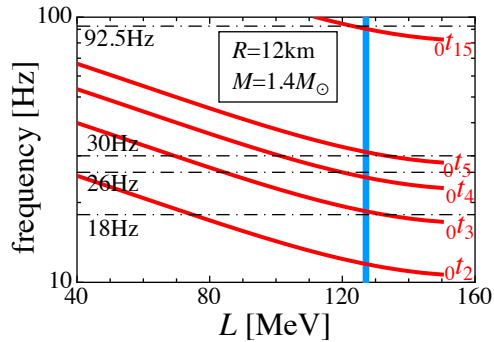} 
\end{center}
\caption{
(Color online) Comparison of the calculated frequencies of torsional 
oscillations of a star having $M=1.4M_\odot$ and $R=12$ km (solid lines) 
with the QPO frequencies observed in SGR 1806-20 
(dot-dashed lines), where we adopt the Chamel data for $N_s/N_d$.
The vertical line corresponds to
the value of $L$ that is consistent with the observations.
}
\label{fig:fit-M14R12}
\end{figure}

Instead of just considering $L_{\rm min}$, we now proceed to obtain a more
stringent constraint on $L$ by fitting the predicted frequencies of 
fundamental torsional oscillations with different values of $\ell$ to the 
low-lying QPO frequencies observed in SGRs.  To this end, we use the 
values of $N_s/N_d$ derived by \cite{Chamel2012}.  In
the present analysis, we focus on the observed QPO frequencies lower than 100 
Hz, i.e., 18, 26, 30, and 92.5 Hz in SGR 1806-20 and 28, 54, and 84 Hz in SGR 
1900+14 \citep{WS2006}. In fact, the even higher observed 
frequencies would be easier to explain in terms of multipolar 
fundamental and overtone frequencies
\footnote{
The possibility to explain the higher QPO frequencies with the shear oscillations in hadron-quark mixed phase in the core of neutron stars is also suggested in \cite{SMT2012}.
}. Because of the small interval between 
the observed frequencies 26 and 30 Hz in SGR 1806-20, it is more 
difficult to explain the QPO frequencies observed in SGR 1806-20 than
those in SGR 1900+14 \citep{Sotani2007}. 
If one identifies the lowest 
frequency in SGR 1806-20 (18 Hz) as the fundamental torsional oscillation with 
$\ell=3$ as in \cite{Sotani2011b}, one can manage to explain 26, 
30, and 92.5 Hz in terms of those with $\ell=4$, 5, and 15.  In the case 
of the typical neutron star model with $M=1.4M_\odot$ and $R=12$ km,
we compare the predicted frequencies with the observed ones as shown in 
Fig.\ \ref{fig:fit-M14R12}.  One can observe from this figure that the best 
value of $L$ to reproduce the observed frequencies is $L=127.1$ MeV, 
where the calculated frequencies, 18.5 Hz ($\ell=3$), 24.9 Hz ($\ell=4$), 
31.0 Hz ($\ell=5$), and 90.3 Hz ($\ell=15$), are within less than $5\%$ 
deviations from the observations.

\begin{figure}
\begin{center}
\includegraphics[scale=0.5]{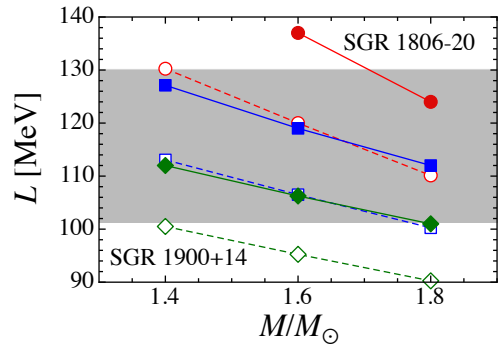} 
\end{center}
\caption{
(Color online) Values of $L$ at which the calculated frequencies 
of torsional oscillations agree best with the QPO frequencies observed 
in SGR 1806-20 (filled symbols with solid lines) and in SGR 1900+14 
(open symbols with broken lines), where the lines are just eye guides. The circles, squares, and 
diamonds correspond to the stellar models with $R=10, 12$, and 14 km, respectively.
The gray region denotes the allowed values of 
$L$ that can be obtained from both of the QPO observations in SGR 
1806-20 and SGR 1900+14 by assuming that the corresponding neutron stars
have mass and radius in the range $1.4\le M/M_\odot\le 1.8$ and 
$10$ km $\le R \le 14$ km.
}
\label{fig:Lfit}
\end{figure}

Let us now extend the analysis to different stellar models and to SGR
1900+14.  We find that the QPOs observed in SGR 1806-20 can be 
explained in terms of the eigenfrequencies of the same $\ell$ within 
similar deviations even for different stellar models except for the case 
with $M=1.4M_\odot$ and $R=10$ km.   The obtained best values of $L$ to 
explain the observations are shown in Fig.\ \ref{fig:Lfit} by solid 
lines and filled symbols.  On the other hand, the low-lying QPOs 
observed in SGR 1900+14 can be similarly explained in terms of 
the fundamental torsional oscillations with $\ell=4$, 8, and 13; the obtained 
best values of $L$ are shown in Fig.\ \ref{fig:Lfit} by broken 
lines and open symbols. As a result, the allowed region of $L$
where the QPO frequencies observed in SGR 1806-20 and in SGR 1900+14 are 
reproducible simultaneously lies in the range 100 MeV $\lsim L\lsim 130$ MeV, 
as long as the oscillating neutron stars have mass and radius ranging
$1.4\le M/M_\odot\le 1.8$ and 10 km $\le R\le$ 14 km.  It is interesting
to compare this constraint with various experimental constraints on $L$ 
(See, e.g., \cite{LCK2008,Tsang2009}), which have yet to converge but seemingly favor smaller $L$.
We remark that as in \cite{SW2009}, one can also explain the three QPOs
observed in SGR 1900+14 in terms of the $\ell = 3$, 6, 9 oscillations
with comparable or even better accuracy, but the optimal values of $L$, 
which are typically of order 80 MeV, 
are too low to become 
consistent with those based on the identification of 
the four QPOs in SGR 1806-20 as the $\ell=3$, 4, 5, 15 oscillations.

In summary, we have examined how neutron superfluidity affects the 
fundamental torsional oscillations in the crustal region of a
neutron star for various EOS and stellar models.  Recent calculations of 
the superfluid density allow us to determine the increase in the frequencies
of fundamental torsional oscillations, of which the comparison with 
the QPO frequencies observed in SGRs gives a more stringent constraint
on the parameter $L$ characterizing the density dependence of the symmetry 
energy than the earlier work ignoring the superfluidity \citep{SNIO2012}.  If
the mass and radius of the oscillating neutron stars are observationally
determined, one might be able to justify the assumption that the QPOs come
from the torsional modes and then to fix $L$ with reasonable accuracy.
We remark that there could be another way of identifying the low-lying QPOs, although the 
26 Hz QPO in SGR 1806-20 remains to be clearly identified \citep{SW2009}. 
In these identifications, 
the 18, 30, 92.5 Hz QPOs in SGR 1806-20 correspond to the $\ell= 2$, 3, 10 oscillations, while 
the 28, 54, 84 Hz QPOs in SGR 1900+14 correspond to the $\ell = 3$, 6, 9 oscillations.
By assuming again that the masses and radii of the oscillating neutron stars 
satisfy $1.4\leq M/M_\odot\leq1.8$ and 10 km $\leq R \leq$ 14 km, we obtain a range of $L$
of $\sim60-80$ MeV in which all the six QPO frequencies are reproducible.
It would be interesting to note the difference from the $L$ values of $\sim100-130$ MeV
obtained above and, if the 26 Hz QPO in SGR 1806-20 could be successfully identified
as a different type of  oscillation from the torsional ones, to consider 
which of these two allowed ranges of $L$ is more favored.

We are grateful to  A. W. Steiner for useful discussions.  This work was 
supported in part by Grants-in-Aid for Scientific Research on 
Innovative Areas through No.\ 23105711 and No.\ 24105008 provided by 
MEXT and in part by Grant-in-Aid for Young Scientists (B) through 
No.\ 24740177 and for Research Activity Start-up through No.\ 23840038 
provided by JSPS.




\begin{thebibliography}{999}

\bibitem[\protect\citeauthoryear{Andersson \& Kokkotas}{1996}]{AK1996}
   Andersson N., Kokkotas K. D., 1996, Phys. Rev. Lett., 677, 4134

\bibitem[\protect\citeauthoryear{Andersson, Glampedakis, \& Samuelsson}{2009}]{AGS2009}
   Andersson N., Glampedakis K., Samuelsson L., 2009, MNRAS, 396, 894

\bibitem[\protect\citeauthoryear{Chamel}{2012}]{Chamel2012}
   Chamel N., 2012, Phys. Rev. C, 85, 035801

\bibitem[\protect\citeauthoryear{Colaiuda \& Kokkotas}{2011}]{CK2011}
   Colaiuda A., Kokkotas K. D., 2011, MNRAS, 414, 3014

\bibitem[\protect\citeauthoryear{Demorest et al.}{2010}]{2Msun}  
   Demorest P. B., Pennucci T., Ransom S. M., Roberts M. S. E., Hessels J. W. T., 2010, Nature (London), 467, 1081

\bibitem[\protect\citeauthoryear{Gabler et al.}{2011}]{GCFMS2011}
   Gabler M., Cerd\'{a}-Dur\'{a}n P., Font J. A., M\"{u}ller E., Stergioulas N., 2011, MNRAS, 410, L37

\bibitem[\protect\citeauthoryear{Gabler et al.}{2012a}]{GCFMS2012a}
   Gabler M., Cerd\'{a}-Dur\'{a}n P., Stergioulas N., Font J. A., M\"{u}ller E., 2012a, MNRAS, 421, 2054
   
\bibitem[\protect\citeauthoryear{Gabler et al.}{2012b}]{GCFMS2012b}
   Gabler M., Cerd\'{a}-Dur\'{a}n P., Font J. A., Muller E., Stergioulas N., 2012b, preprint (arXiv:1208.6443)

\bibitem[\protect\citeauthoryear{Gearheart et al.}{2011}]{GNHL2011}
   Gearheart M., Newton W. G., Hooker J., Li B. A., 2011, MNRAS, 418, 2343

\bibitem[\protect\citeauthoryear{Hurley et al.}{1999}]{H1999}
   Hurley K. et al., 1999, Nature, 397, L41

\bibitem[\protect\citeauthoryear{Iida \& Sato}{1997}]{IS1997}
   Iida K., Sato K., 1997, ApJ, 477, 294

\bibitem[\protect\citeauthoryear{Iida \& Baym}{2002}]{IB-II}
   Iida K., Baym G., 2002, Phys. Rev. D, 65, 014022

\bibitem[\protect\citeauthoryear{Kouveliotou et al.}{1998}]{K1998}
   Kouveliotou C. et al., 1998, Nature, 393, L235

\bibitem[\protect\citeauthoryear{Lattimer}{1981}]{L1981}
   Lattimer J. M., 1981, Annu. Rev. Nucl. Part. Sci., 31, 337

\bibitem[\protect\citeauthoryear{Lee}{2007}]{Lee2007}
   Lee U., 2007, MNRAS, 374, 1015

\bibitem[\protect\citeauthoryear{Levin}{2006}]{Levin2006}
   Levin Y., 2006, MNRAS, 368, L35

\bibitem[\protect\citeauthoryear{Li, Chen, \& Ko}{2008}]{LCK2008} 
   Li B. A., Chen C. W., Ko C. M., 2008, Phys. Rep., 464, 113
   
\bibitem[\protect\citeauthoryear{Lorenz et al.}{1993}]{LRP1993}
   Lorenz C. P., Ravenhall D. G., Pethick C. J., 1993, Phys. Rev. Lett., 70, 379

\bibitem[\protect\citeauthoryear{Ogata \& Ichimaru}{1990}]{OI1990}
   Ogata S., Ichimaru S., 1990, Phys. Rev. A, 42, 4867

\bibitem[\protect\citeauthoryear{Oyamatsu}{1993}]{O1993}
   Oyamatsu K., 1993, Nucl. Phys. A, 561, 431

\bibitem[\protect\citeauthoryear{Oyamatsu \& Iida}{2003}]{OI2003}
   Oyamatsu K., Iida K., 2003, Prog. Theor. Phys., 109, 631

\bibitem[\protect\citeauthoryear{Oyamatsu \& Iida}{2007}]{OI2007}
   Oyamatsu K., Iida K., 2007, Phys. Rev. C, 75, 015801


\bibitem[\protect\citeauthoryear{Passamonti \& Andersson}{2012}]{PA2012}
   Passamonti A., Andersson N., 2012, MNRAS, 419, 638

\bibitem[\protect\citeauthoryear{Pethick \& Potekhin}{1998}]{PP1998}
   Pethick C. J., Potekhin A. Y., 1998, Phys. Lett. B, 427, 7

\bibitem[\protect\citeauthoryear{Pethick, Chamel, \& Reddy}{2010}]{PCR2010}
   Pethick C. J., Chamel N., Reddy S., 2010, Prog. Theor. Phys. Suppl., 186, 9

\bibitem[\protect\citeauthoryear{Samuelsson \& Andersson}{2007}]{SA2007}
   Samuelsson L., Andersson N., 2007, MNRAS, 374, 256

\bibitem[\protect\citeauthoryear{Samuelsson \& Andersson}{2009}]{SA2009}
   Samuelsson L., Andersson N., 2009, Class. Quant. Gravity, 26, 155016

\bibitem[\protect\citeauthoryear{Sauls}{1989}]{Sauls}
   Sauls J. A., in {\it Timing Neutron Stars}, edited by \"{O}gelman H., van den Heuvel E.P.J. (Kluwer, Dortrecht, 1989), P. 457.

\bibitem[\protect\citeauthoryear{Schumaker \& Thorne}{1983}]{ST1983}
   Schumaker B. L., Thorne K. S., 1983, MNRAS, 203, 457


\bibitem[\protect\citeauthoryear{Sotani, Tominaga, \& Maeda}{2001}]{Sotani2001}
   Sotani H., Tominaga K., Maeda K. I., 2001, Phys. Rev. D, 65, 024010

\bibitem[\protect\citeauthoryear{Sotani, Kohri, \& Harada}{2004}]{Sotani2004}
   Sotani H., Kohri K., Harada T., 2004, Phys. Rev. D, 69, 084008

\bibitem[\protect\citeauthoryear{Sotani, Kokkotas, \& Stergioulas}{2007}]{Sotani2007}
   Sotani H., Kokkotas K.D., N. Stergioulas, 2007, MNRAS, 375, 261

\bibitem[\protect\citeauthoryear{Sotani, Kokkotas, \& Stergioulas}{2008}]{Sotani2008a}
   Sotani H., Kokkotas K. D., Stergioulas N., 2008, MNRAS, 385, L5

\bibitem[\protect\citeauthoryear{Sotani, Colaiuda, \& Kokkotas}{2008}]{Sotani2008b}
   Sotani H., Colaiuda A., Kokkotas K. D., 2008, MNRAS, 385, 2161

\bibitem[\protect\citeauthoryear{Sotani \& Kokkotas}{2009}]{Sotani2009}
   Sotani H., Kokkotas K. D., 2009, MNRAS, 395, 1163

\bibitem[\protect\citeauthoryear{Sotani}{2011}]{Sotani2011b}
   Sotani H., 2011, MNRAS, 417, L70

\bibitem[\protect\citeauthoryear{Sotani et al.}{2011}]{SYMT2011}
   Sotani H., Yasutake N., Maruyama T., Tatsumi T., 2011, Phys. Rev. D, 83, 024014

\bibitem[\protect\citeauthoryear{Sotani et al.}{2012}]{SNIO2012}
   Sotani H., Nakazato K., Iida K., Oyamatsu K., 2012, Phys. Rev. Lett., 108, 201101

\bibitem[\protect\citeauthoryear{Sotani, Maruyama, \& Tatsumi}{2012}]{SMT2012}
   Sotani H., Maruyama T., Tatsumi T., 2012, preprint (arXiv:1207.4055)

\bibitem[\protect\citeauthoryear{Steiner \& Watts}{2009}]{SW2009}
   Steiner A. W., Watts A. L., 2009, Phys. Rev. Lett., 103, 181101

\bibitem[\protect\citeauthoryear{Strohmayer et al.}{1991}]{SHOII1991}
   Strohmayer T., van Horn H. M., Ogata S., Iyetomi H., Ichimaru S., 1991, ApJ., 375, 679

\bibitem[\protect\citeauthoryear{Tsang et al.}{2009}]{Tsang2009} 
   Tsang M. B. {\it et al.}, 2009, Phys. Rev. Lett., 102, 122701

\bibitem[\protect\citeauthoryear{van Horn \& Epstein}{1990}]{vanhorn90}
   van Horn H. M., Epstein R. I., 1990, Bull. American Astron. Soc., 22, 748

\bibitem[\protect\citeauthoryear{Watts \& Strohmayer}{2006}]{WS2006}
   Watts A. L., Strohmayer T. E., 2006, Adv. Space Res., 40, 1446






\end{thebibliography}
\end{document}